\documentclass[preprint,authoryear,11pt,5p,times]{elsarticle}
\usepackage{times}
\usepackage{bm}
\usepackage{verbatim, amsmath, amsfonts, amssymb}
\usepackage[normalem]{ulem} 
\usepackage{latexsym}%
\usepackage{graphicx}%
\usepackage{booktabs}
\usepackage[breaklinks=true,colorlinks=true]{hyperref}
\usepackage{listings}
\usepackage{cleveref} 
\usepackage{url}
\usepackage[multiple]{footmisc}
\usepackage[table]{xcolor}
\usepackage[many]{tcolorbox}
\usepackage[T1]{fontenc}
\usepackage[utf8]{inputenc}

\definecolor{tealblue}{rgb}{0.21, 0.46, 0.53}

\journal{Astronomy And Computing}
\newcommand\aap{A\&A}                
\newcommand\aaps{A\&AS}              
\newcommand\aj{AJ}                   
\newcommand\ao{Appl. Opt.}           
\newcommand\apj{ApJ}                 
\newcommand\apjl{ApJ}                
\newcommand\apjs{ApJS}               
\newcommand\araa{ARA\&A}             
\newcommand\mnras{MNRAS}             
\newcommand\pre{Phys. Rev.~E}        
\newcommand\pasp{PASP}               

\DeclareRobustCommand{\ion}[2]{%
\relax\ifmmode
\ifx\testbx\f@series
{\mathbf{#1\,\mathsc{#2}}}\else
{\mathrm{#1\,\mathsc{#2}}}\fi
\else\textup{#1\,{\mdseries\textsc{#2}}}%
\fi}

\newcommand\umap{{\sc umap~}}

\begin{document}

\begin{frontmatter}

\title{A graph-based spectral classification of Type II supernovae}

\author[1]{Rafael S. de Souza\corref{mycorrespondingauthor}}
\cortext[mycorrespondingauthor]{Corresponding author}
\ead{r.da-silva-de-souza@herts.ac.uk}
\author[2,3]{Stephen Thorp}
\author[4,5]{Llu\'is Galbany}
\author[6]{Emille E. O. Ishida}
\author[7]{Santiago González-Gaitán}
\author[8]{Morgan A. Schmitz}
\author[7,9]{Alberto Krone-Martins}
\author[10]{Christina Peters}
\author{for the COIN collaboration}

\address[1]{Centre for Astrophysics Research, University of Hertfordshire, College Lane, Hatfield, AL10~9AB, UK
}
\address[2]{Institute of Astronomy, University of Cambridge, Madingley Road, Cambridge, CB3 0HA, UK}
\address[3]{The Oskar Klein Centre, Department of Physics, Stockholm University, AlbaNova University Centre, SE 106 91 Stockholm, Sweden}
\address[4]{Institute of Space Sciences (ICE, CSIC), Campus UAB, Carrer de Can Magrans, s$/$n,
E-08193 Barcelona, Spain
}
\address[5]{Institut d'Estudis Espacials de Catalunya (IEEC), E-08034 Barcelona, Spain
}
\address[6]{Universit\'e Clermont Auvergne, CNRS$/$IN2P3, LPC, F-63000 Clermont-Ferrand, France}
\address[7]{CENTRA-Centro de Astrofísica e Gravita\c{c}ão and Departamento de Física, Instituto Superior Técnico, Universidade de Lisboa, Avenida Rovisco Pais, 1049-001 Lisboa, Portugal
}
\address[8]{Universit\'e C\^{o}te d’Azur, Observatoire de la C\^{o}te d’Azur, CNRS, Laboratoire Lagrange, Bd de l’Observatoire, CS 34229, 06304 Nice Cedex 4, France.
}
\address[9]{Donald Bren School of Information and Computer Sciences, University of California, Irvine, CA 92697, USA
}
\address[10]{Department of Computer and Information Sciences, University of Delaware, Newark, DE, USA
}

\begin{abstract}
Given the ever-increasing number of time-domain astronomical surveys, employing robust, interpretative, and automated data-driven classification schemes is pivotal. Based on graph theory, we present new data-driven classification heuristics for spectral data.
A spectral classification scheme of Type II supernovae (SNe II) is proposed based on the phase relative to the maximum light in the $V$ band and the end of the plateau phase. 
We utilize a compiled optical data set that comprises 145 SNe and 1595 optical spectra in 4000-9000 \AA.
Our classification method naturally identifies outliers and arranges the different SNe in terms of their major spectral features. We compare our approach to the off-the-shelf \umap manifold learning and show that both strategies are consistent with a continuous variation of spectral types rather than discrete families. The automated classification naturally reflects the fast evolution of Type II SNe around the maximum light while showcasing their homogeneity close to the end of the plateau phase. The scheme we develop could be more widely applicable to unsupervised time series classification or characterisation of other functional data.
\end{abstract}


\end{frontmatter}


\section{Introduction}

Classification systems have a long history, dating back to ancient times. One of the oldest documented examples is a pharmacopoeia called "The Divine Farmer's Materia Medica", which dates from around 200 AD \citep{yang1998divine}. However, it was only centuries later that the foundations of modern taxonomy were laid by the Linnaean classification of organisms \citep{linnaeus1758} and Darwin's theory of common descent \citep{darwin1859origin}. In astronomy, two well-known examples of classification systems are the Morgan-Keenan system of stellar spectral classification \citep{MK1973} and the Hertzsprung-Russell diagram \citep[e.g.][and references therein]{Chiosi1992}, which organise stars according to their luminosity, effective temperature, and evolutionary stage. With the advent of large-scale surveys, the diversity of observed objects has considerably increased, making the construction of a coherent taxonomy that embraces this plurality a particularly challenging task.

One essential tool employed in such tasks is graph networks. Graph theory originated in Königsberg, the capital of Eastern Prussia, in 1735, when Leonard Euler offered a rigorous mathematical proof that there is no path to walk across all seven bridges over the Pregel river without crossing one bridge at least twice. This solution is the first documented real-world problem solved by a graph \citep{Rob2012}.

More than a historical anecdote, this episode also elucidates that some problems can be solved by changing the paradigm we approach them. Graphs have been previously used in astronomy to visualise multivariate tabular data \citep{DESOUZA2015100} and have gained popularity in diverse fields, including ecology \citep{Farage2021}, neural sciences \citep{Bessadok2021,LI2021102233}, and computer vision \citep{Vasudevan2022}. They are particularly suitable for simplifying and highlighting hidden data structures and associations in multidimensional and heterogeneous datasets.

A notable example of a problem involving heterogeneous data is the challenge of building a coherent classification system for supernovae explosions.  Supernovae classification relies on spectra and has traditionally been divided into two main classes: Type I SNe (SNe I) -- those lacking any hydrogen emission lines in their spectra -- and Type II SNe (SNe II) -- those with hydrogen emission \citep{minkowski1941}. SNe II are a diverse group exhibiting many photometric and spectroscopic features, including a broad range of brightness, visual decline rates, photospheric phase duration, and spectral line features.

In particular, Type II SNe have been historically divided into two groups based on their light-curve properties: IIP -- sharp rise to the peak after $\sim$ 5-15 days of the explosion, followed by a plateau phase of nearly 70-120 days, and  IIL --  shorter cooling phase and faster luminosity decline rate \citep{GalYam2017}. Conversely, other classes are defined based on their spectroscopic properties:  IIn \citep{Schlegel1990} --  long-lasting narrow  Balmer emission lines -- and IIb \citep{Filippenko1988} -- transitional objects 
that begin their evolution similarly to SNe II and evolve towards a state with no $\mathrm{H}_\alpha$ but prominent [\ion{Ca}{ii}] and [\ion{O}{i}] lines (typical of SNe Ib ). It has been suggested that SNe IIb are physically distinct from other SNe II \citep[e.g.][]{Pessi2019}. An additional photometric subclass, namely SN 1987A-like SNe, displays a  peculiar long rise to maximum \citep[e.g.,][]{McCray2017}, with spectra similar to SNe IIP \citep{Modjaz2019Nat}.  The connection between different SN II subtypes and their physical origin has been discussed extensively. Direct identification of SN IIP progenitors has suggested a red supergiant (RSG) origin (with mass $\sim8-17\mathrm{M}_\odot$; see e.g.\ \citealp{Smartt2009}). Candidate SN IIL progenitors have been more difficult to identify, although tentative identifications have suggested that these may be more massive \citep[e.g.][]{EliasRosa2010, EliasRosa2011}. Some photometric analyses have suggested that SNe IIP and IIL are fully distinct with different physical mechanisms \citep[e.g. red supergiants vs.\ magnetars; see][]{Arcavi2012}, whereas others have pointed to a single continuous family \citep[e.g.][]{Anderson2014ApJ,Gutierrez2014,Galbany2016AJ, Valenti2016} with observational properties driven by the scale and density of the hydrodgen envelopes on the progenitors. Furthermore, in recent years, a large fraction of SNe~II have been shown to present circumstellar material ejected before explosion -- or extended atmospheres in the progenitor RSGs -- explaining the fast rise in the light-curves \citep[e.g.][]{Gonzalez-Gaitan2015MNRAS,Morozova2017,Forster2018} and the clear narrow emission signatures in early spectra \citep[e.g.][]{Yaron2017}.

Traditional classification methods include template matching \citep{Howell2005,Blondin2007, Harutyunyan2008}, and similarity of specific spectral features \citep{Sun2017,Prentice2017}. Re-classification schemes have been proposed because of their wide spectral diversity, photometric features, and the limitations of the template-matching approach, given the lack of enough representative spectra.  \citet{Ransome2021} proposed a new classification scheme for Type IIn solely based on a detectable narrow feature in the $\mathrm{H}_\alpha$ profile, whereas \citet{Anderson2014ApJ} suggested using the post-maximum decline rate to classify SNe~II. Alternatively, data-driven approaches have been developed in previous years. 

Noteworthy examples of spectral classification methods include \citet{Williamson2019}, who utilized principal component analysis (PCA) for categorizing a sample of 160 stripped-envelope core-collapse supernovae. \citet{Chen2018} employed functional PCA to extract spectral features of SNe IIP/IIL, which were then classified using Support Vector Machine (SVM) and Artificial Neural Network (ANN). For a broader scope of supernova types, \citet{Muthukrishna2019} developed \texttt{dash}, a deep-learning-based model trained on over 4,000 classified spectra for determining a supernova spectrum's type, age, redshift, and host galaxy component. More recently, \citet{Bengyat2022} applied an unsupervised Random Forest algorithm to characterize supernovae based on their spectral-temporal distribution.

Supernova classification often involves both the separation of distinct classes (e.g., SNe Ia vs. core-collapse) and the characterization of potentially continuous subfamilies (e.g., SNe II), where the full continuity of a class may not be known a priori. This work introduces a novel approach that can efficiently handle a mixture of discrete and continuous classes. By employing graph networks for unsupervised learning, it defines a taxonomic landscape of SNe II (SNe IIP and SNe IIL), resulting in a data-driven classification scheme that can rapidly update with new data.

Our heuristics rely on a hierarchical classification of the SNe II spectra based on similarity, conveying an intuitive graph-based visualisation. To account for the time evolution of SNe spectra, we apply our analysis over two reference phases of spectral evolution, around maximum light and the end of the plateau phase. Additionally, we compare our methodology with more complex manifold-learning-based classification and find they lead to similar results, favouring a continuous spectral variation instead of compact families. 

The rest of the paper is organised as follows. 
\autoref{sec:data} describes our data compilation and pre-processing. \autoref{sec:method} describes the heuristics behind the automated spectral taxonomy scheme. \autoref{sec:groups} displays the spectral classification of Type II SNe for two different phases.  \autoref{sec:concl} finalizes with our conclusions. 

\section{Data description}
\label{sec:data}

In this section, we describe the spectral and photometric dataset and pre-processing used in this work. 

\subsection{Sample selection}
We begin by compiling publicly available data for 568 photometric SNe~II, 415 of which also have publicly available spectroscopic information. Our final sample of photometric light-curves primarily includes data from large surveys, including the Cerro-Tololo Survey, C\'alan-Tololo Survey, Supernova Optical and Infrared Survey, and Carnegie Type II Supernova Survey \citep{Galbany2016AJ};  Lick Observatory Supernova Search \citep{Leaman2011, Faran2014, deJaeger2019}; Center for Astrophysics Supernova Program \citep{Hicken2017}; Carnegie Supernova Project \citep{Hamuy2006}; the Las Cumbres Observatory \citep{Valenti2016}; the Sloan Digital Sky Survey-II Supernova Survey \citep{Dandrea2010}; the \textit{Swift} UV-Optical Telescope \citep{Pritchard2014}. We also use photometric data for individual objects published elsewhere in the literature \citep{Hamuy1990, Benetti1991, Benetti1994, Schmidt1993, Schmidt1994, Blanton1995, Cappellaro1995, Clocchiatti1996, Hamuy2001, Leonard2002_99em, Leonard2002_99gi, Pastorello2004, Pastorello2005, Pastorello2009, Pastorello2012, Chugai2005, Hendry2005, Hendry2006, Sahu2006, Tsvetkov2006_01X, Tsvetkov2008_04ek, Tsvetkov2006, Tsvetkov2008_04dj, Vinko2006, Vinko2009, Zhang2006, Misra2007, Dessart2008, Krisciunas2009, Maguire2010, Fraser2011, Inserra2011, Inserra2012, Inserra2013,  Kleiser2011, Roy2011, VanDyk2012, Bose2013, Bose2015_13ej, Bose2016, Gandhi2013, Tomasella2013, Tomasella2018, Dallora2014, Richmond2014, Spiro2014, Zhang2014, Barbarino2015, Huang2015, Jerkstrand2015, Kangas2016}.

Our spectroscopic data are predominantly compiled from the Carnegie Type-II Supernova Survey and Carnegie Supernova Project \citep{Gutierrez2017} or taken from the Weizmann Interactive Supernova Data Repository \citep{Yaron2012} and Asiago Supernova Catalogue \citep{Barbon1989, Barbon1999}. We also make use of data from the Public ESO Spectroscopic Survey of Transient Objects \citep{Smartt2015}, Lick Observatory Supernova Search \citep{deJaeger2019}, Center for Astrophysics Supernova Program \citep{Modjaz2014, Hicken2017}, and elsewhere in the literature including \citet{Inserra2012, Faran2014, Zhang2014}.

We require a successful estimate of $V$-band maximum date via a polynomial fit to the light curve \citep[as in][]{Galbany2016AJ}. This step narrows the sample to 186 SNe, of which 180 have at least two observed spectra. Starting from this selection of 180 SNe~II, we carry out our preprocessing (see \autoref{sec:preprocessing}) to homogenise the data, leading to a final sample of 145 SNe~II. 

\subsection{Pre-processing}
\label{sec:preprocessing}

We create a flux-calibrated spectroscopic sequence sample by partially following the scheme of \citet{vincenzi2019}\footnote{\url{https://github.com/maria-vincenzi/PyCoCo_templates}}. The first step is to perform a Gaussian process (GP) regression \citep[see, e.g.,][]{rasmussen2006} on the photometric light curves, allowing one to estimate fluxes at the times of the observed spectra. We choose not to adopt the physically motivated extrapolation of \citet{vincenzi2019} and strictly interpolate our light curves.  Recently, \citet{Stevance2022} highlighted some challenges associated with GP regression on SN II light curves. Our fits are carried out using a Mat\'ern-$3/2$ kernel, identified by \citealp{Stevance2022} as reasonably fitting the fast transitions in SN IIP light curves (something we also find to be true in practice). We have avoided extrapolating or estimating photometric parameters and have visually inspected for overfitting. We have found that the GP regression method is robust enough for interpolation, which is necessary for flux calibration.

After smoothing the observed spectra with a Savitzky-Golay filter of window size 100~\AA\ \citep{savitzky1964}, we synthesize the fluxes in the passbands in which photometry is available. We then compare those synthesized fluxes to fluxes interpolated from the light curves, allowing us to estimate smooth mangling functions to flux-calibrate the spectra. This step uses a second GP regression in the wavelength direction, with a fixed scale of 300~\AA. The light curves and flux-calibrated spectra are then jointly modelled as a two-dimensional GP in time and wavelength, under the approximation that the light curve points are observations of this surface at the central wavelengths of their passbands. We use the same Mat\'ern-$3/2$ kernel as \citet{vincenzi2019}, with a fixed time and wavelength-scale of 30~days and 100~\AA. The procedure enables all spectra and photometry for a given supernova to provide a principled extrapolation of the observed spectra to unobserved wavelengths. 

\begin{figure}
    \centering
    \includegraphics[width=0.975\linewidth]{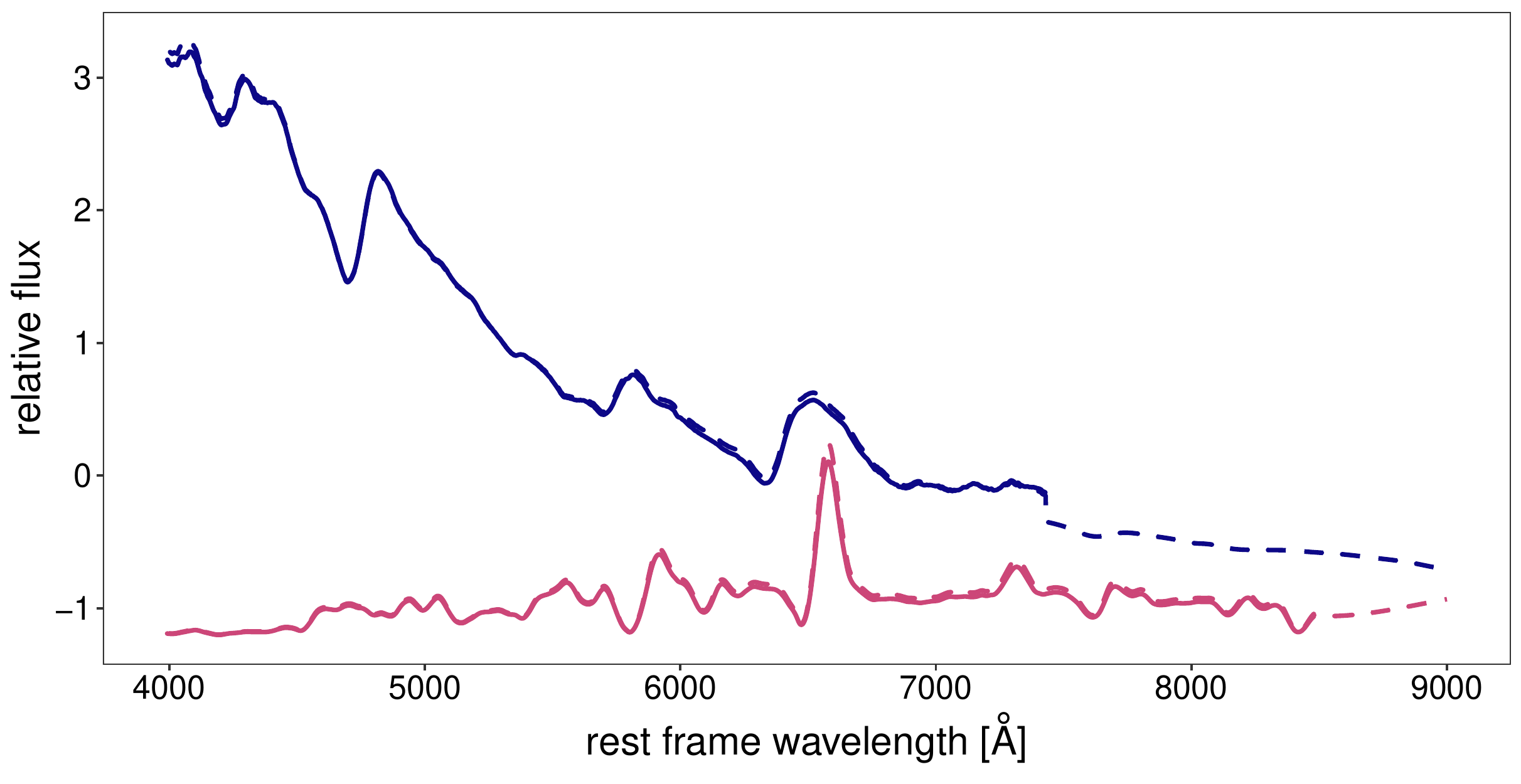}
    \caption{Smoothed original (solid lines) and extended estimated (dashed lines) spectra at maximum light (blue spectra) and at end of plateau phase (red spectra) for SN1999em.}
    \label{fig:spec_rec}
\end{figure}

We do not interpolate the flux surface to unobserved times, as the GP can over-smooth spectral features if this interpolation strays far from the epochs of real spectra. Additionally, we do not adopt the priors of \cite{vincenzi2019}. Instead, we rely solely on the data to justify extrapolation on a comparable wavelength range, which we carry out within the 4000--9000~\AA range.  Finally, the second round of mangling is performed on the extended spectra, yielding a set of flux-calibrated spectra on a consistent wavelength grid. We correct for Milky Way extinction but do not attempt to correct for host galaxy dust. A sample of 145 SNe~II  with a total of 1595 spectra\footnote{For the sample of 145 SNe II, the median number of spectra per SNe is eight. The most well-covered supernova (SN~2012ec) has 68 spectra. There are 38 SNe with more than $1595/145=11$ spectra. The median spectrum epoch is 31.25~days past the estimated time of maximum light, and 42 ~days before the end of the plateau. The interquartile range of our spectral epochs spans 9.12 to 71.79~days post-maximum. The 90th percentile falls at 135.79~days after maximum-light. Around 10 (13) per cent of our spectra are pre-maximum (post-plateau), with 62 (49) supernovae having a pre-maximum (post-plateau) spectrum.} can be successfully pre-processed through the full pipeline. For those that could not be pre-processed in this way, there was at least one stage where a reliable GP regression was not possible. \autoref{fig:spec_rec} shows an example of the process for SN1999em. The solid lines represent the original spectra after Savitzky-Golay smoothing and the first round of mangling. The dashed line represents the final estimated spectra for maximum light and the end of the plateau phase. 

Having mangled and extended the spectra, we sorted them into bins by rest-frame epoch. Specifically, we select two ``target'' epochs of $\pm$ 3 rest-frame days around maximum light and $\pm$ 5 days around the end of the plateau phase\footnote{The time after the plateau or linear decline phase in Type II SNe represents a transition between the photospheric shock-powered hydrogen recombination phase and the start of a radioactive decay-powered $^{56}$Co $\rightarrow$ $^{56}$Fe linear decline phase.}. These two reference epochs are obtained in different ways: the maximum is calculated with polynomial fits in $V$-band, whereas the end of the plateau is obtained by fitting a Fermi-Dirac step function (see e.g. \citealt{Olivares08}). These two reference epochs are thus independent of each other.
We then select the closest spectra falling into one of these bins for each supernova. This procedure yields a sample of 65 SN spectra around maximum light and 27 spectra around the end of the plateau phase.  The remainder of the 145 SNe do not have any spectra sufficiently close to our chosen target epochs.

\section{Graph-based clustering}
\label{sec:method}

This section conveys a brief overview of the methods employed in our work. The steps include the construction of a pairwise dissimilarity matrix from all SNe spectra in a given phase, a minimum spanning tree algorithm ({\tt MST}) to simplify the structure, and a graph community detection to characterize potential groups of similar spectra and outliers. 

\subsection{Dissimilarity Matrix}
\label{sec:dist}

Estimating the proximity between two datasets in high-dimensional space is inherently difficult due to the curse of dimensionality. This is because the performance of similarity indexing structures in high dimensions degrades rapidly. Therefore, the choice of a distance metric is crucial, and it is not always straightforward. Previous studies have shown that for a wide variety of problems, the $\ell_1$ (Manhattan) norm performs better than the $\ell_2$ (Euclidean) norm  \citep{Aggarwal2001}.

The $\ell_p$-norm is given by:
\begin{equation}
L_p(x_i,y_i) = \left(\sum_{i=1}^d \|x_i - y_i\|^p\right)^{\frac{1}{p}}, 
\label{eq:l1}
\end{equation}
Where $x$ and $y$ represent two independent vectors of features, $i$ is the index of the feature and $d$ is the number of features. In our case, the features are the flux values at each wavelength bin, and the number of bins is 501, in the range of 4000 to 9000 \AA. Thus, we choose to use the $\ell_1$-norm to measure the similarity between any two supernova spectra:

\begin{equation}
L_1(x_i,y_i) = \sum_{i=1}^{501} \|x_i - y_i\|.
\end{equation}
Here, $x_i$ and $y_i$ represent the flux values at each wavelength bin $i$ for two different supernova spectra $x$ and $y$. 

\begin{figure*}
    \centering
    \includegraphics[width=0.45\linewidth]{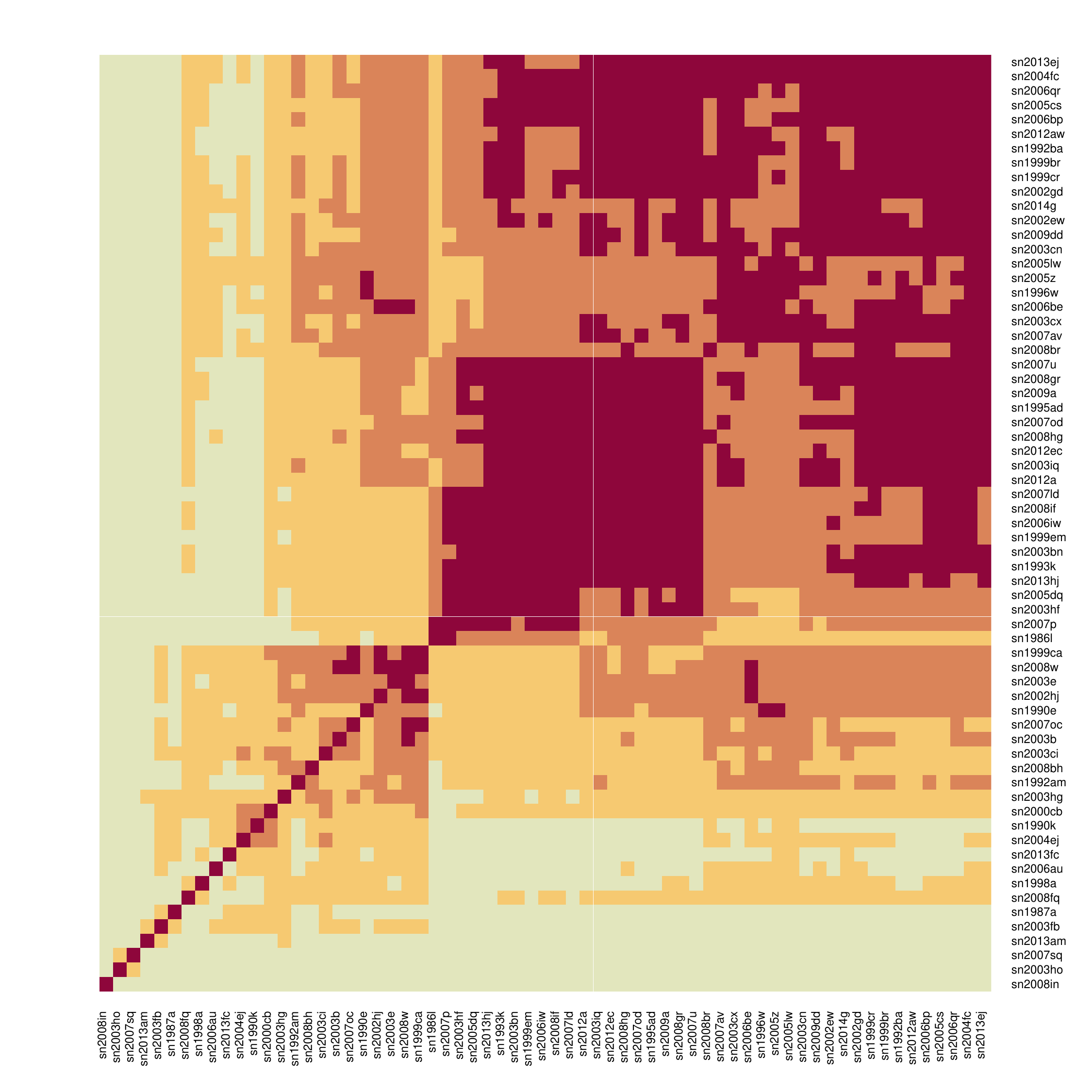}
       \includegraphics[width=0.45\linewidth]{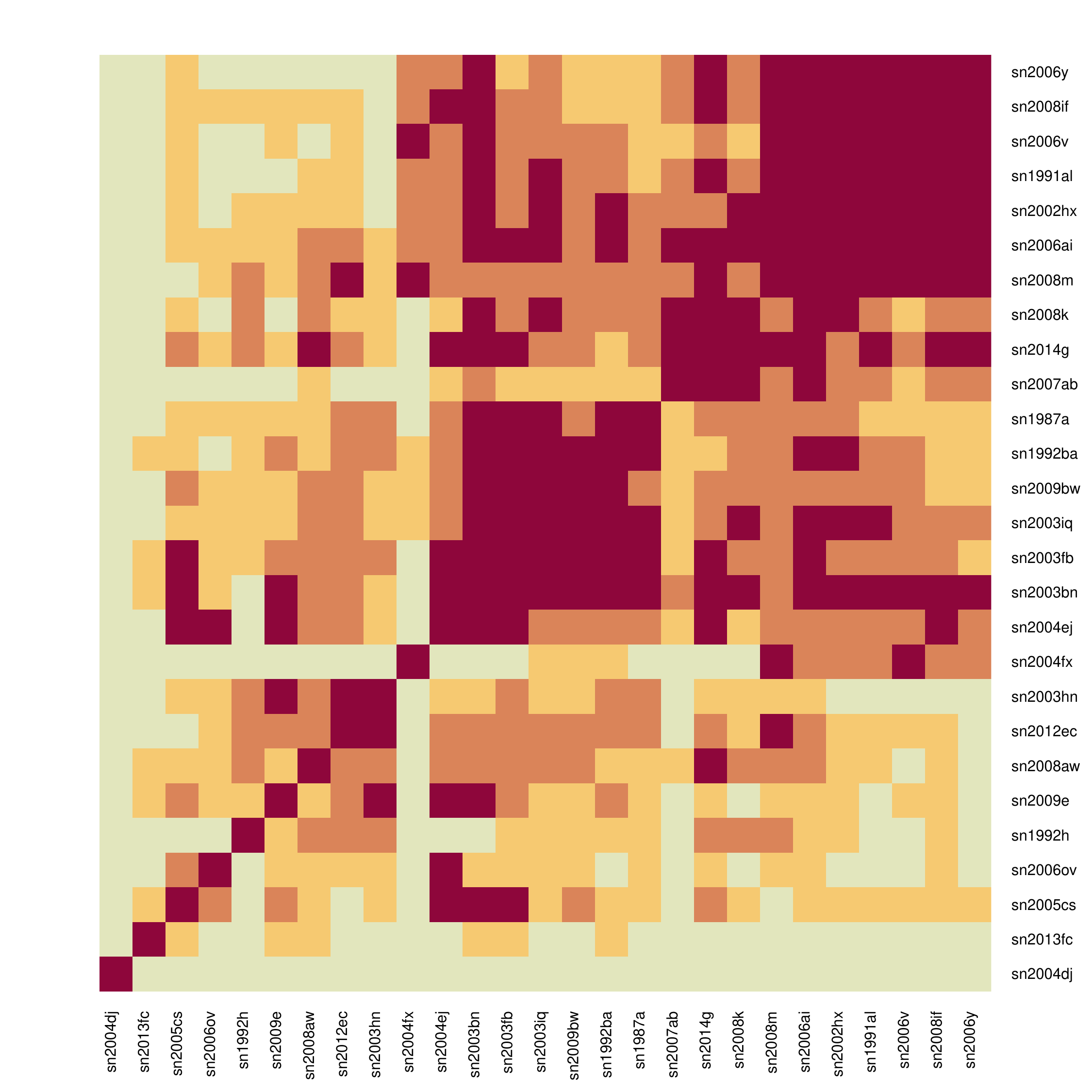}
    \caption{Dissimilarity Matrix of the SNe sample at maximum light (left) and end of plateau phase (right).  The colour scale translates levels of similarity between a given pair of objects, which increases from green to red.}
    \label{fig:heatmap}
\end{figure*}

Given this, the dissimilarity matrix is constructed by calculating the pairwise Manhattan distances between all the spectra in a given phase.
The resulting heatmap of the dissimilarity matrix is shown in \autoref{fig:heatmap} for maximum light and the end of the plateau phase. A simple visual inspection suggests that there is some level of structure in the data. For example, SN2004dj stands out as an outlier at the end of the plateau phase, not correlating strongly with any other supernovae. The next step is to make sense of this information and convey an intuitive visualization. To ensure the robustness of our results, we also conducted a sensitivity analysis using the $\ell_2$-norm and generated a corresponding dissimilarity matrix, which is presented in \ref{app:lnorm}. Although the resulting group arrangement of the supernovae was slightly different from that of the original analysis using the Manhattan distance measure, our subsequent analysis was not qualitatively affected by this change.

\subsection{Connected, Undirected Graphs}

To further understand the relationships between the supernova spectra, we use the concept of a connected, undirected graph. In this approach, each supernova spectrum is represented by a vertex, and the edges connecting these vertices represent the similarities between the spectra, i.e. the closer two SNe the more similar their spectra are. Precisely, the weight or length of each edge corresponds to the degree of similarity between the two supernova spectra that it connects. The closer two supernovae are, the more similar their spectra will be, and the shorter the corresponding edge will be.
A connected, undirected graph is one in which every pair of vertices has a path connecting them. This means at least one sequence of edges connects any two vertices in the graph. Additionally, an undirected graph is one in which the edges do not have a specific direction. This means that if there exists an edge between two vertices, that edge can be traversed in either direction. The edges of a connected, undirected graph connect all the vertices to each other directly or indirectly.
The connected, undirected graph is a useful tool for modelling relationships between objects without a clear discontinuity between them. Additionally, it is also useful for modelling relationships in which the direction of the relationship does not matter. In our case, this approach allows us to understand the similarities between the supernova spectra and to identify potential groups of similar spectra, as well as outliers. The use of a minimum spanning tree algorithm and graph community detection methods can also aid in simplifying the structure and characterizing the relationships in the data.

\subsection{Minimum spanning tree}

Minimum Spanning Tree ({\tt MST}) algorithms have gained widespread popularity as a powerful tool for identifying clusters of heterogeneous data in various fields. Here, the algorithm is applied to  graphs in which the vertices represent SNe spectra and the edges represent the distances between each pair of spectra, as calculated using the $\ell_1$-norm \autoref{eq:l1}.  The MST is a subset of the edges in a connected, undirected graph that spans all the vertices without forming any cycles, while minimizing the total edge weight. In the context of SNe spectra, this weight represents the dissimilarity between each pair of spectra. 

\begin{figure*}
    \centering
     \includegraphics[width=0.45\linewidth]{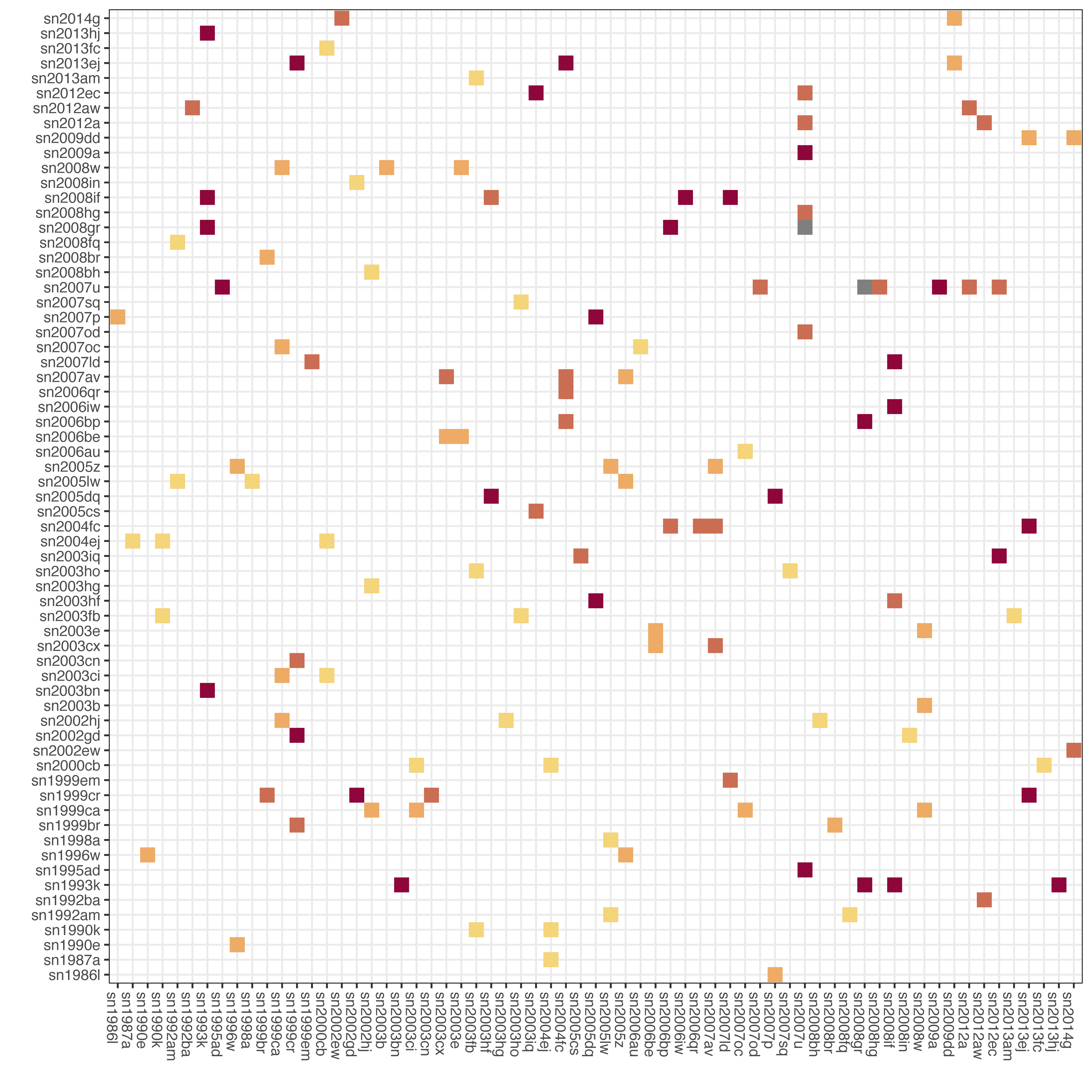}
    \includegraphics[width=0.45\linewidth]{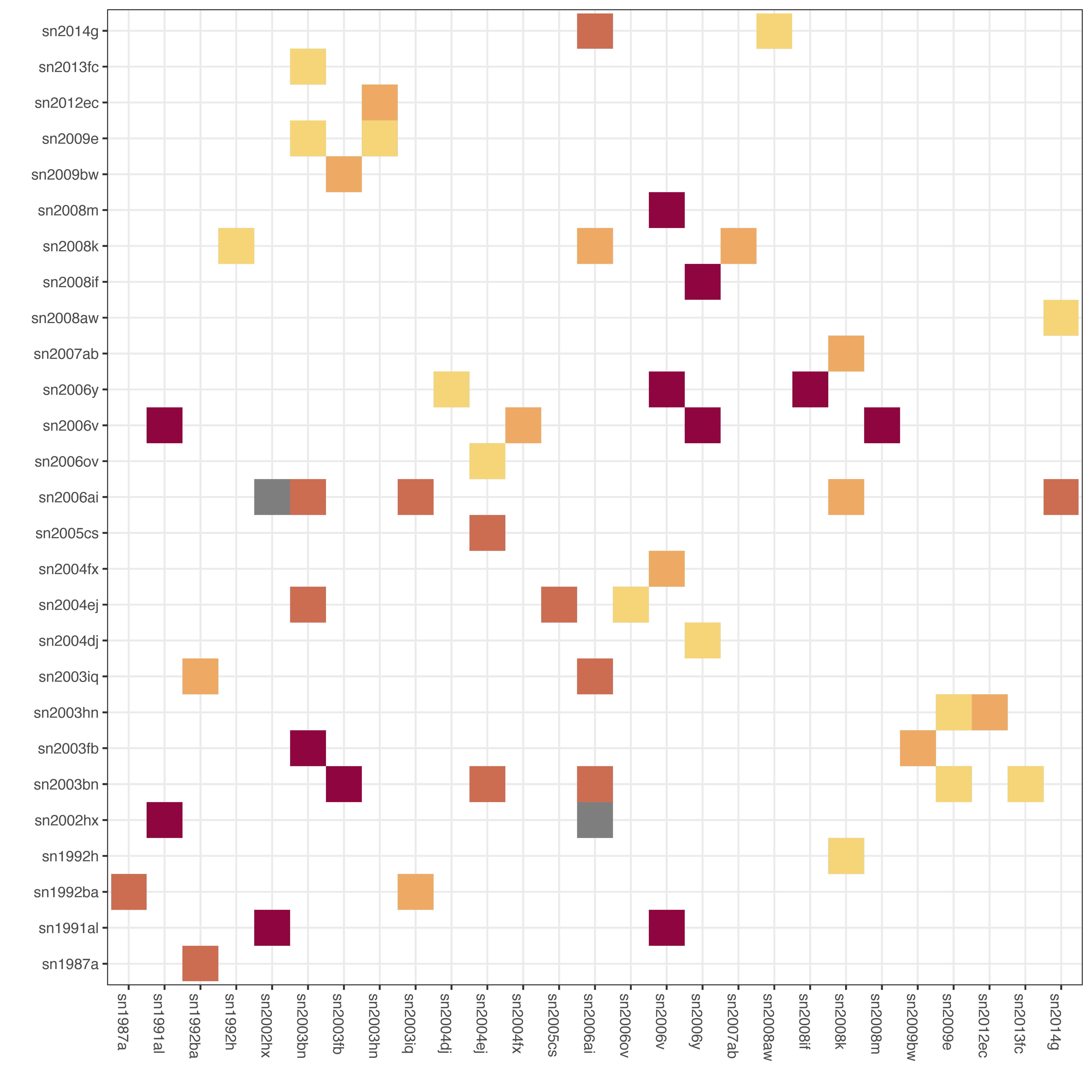}
    \caption{Graph in a matrix form of the SNe sample at maximum light (left) and end of plateau phase (right) after applying the {\tt MST}. The colour scale translates levels of similarity between a given pair of objects, which increases from green to red. }
    \label{fig:heatmap_mst}
\end{figure*}

One of the most widely used and well-known MST algorithms is Prim's algorithm, first proposed by Robert C. Prim in 1957 \citep{Prim1957}. This algorithm starts with an arbitrary root vertex and repeatedly adds the shortest edge that connects a new vertex to the tree. This process is repeated until all vertices are connected, resulting in a multilayered representation of the similarities between all spectra in the data for a given phase. This representation captures global and local associations, allowing for a comprehensive and in-depth data analysis. \autoref{fig:heatmap_mst} shows the matrix representation of the graph after applying the {\tt MST} to the dissimilarity matrix shown in \autoref{fig:heatmap}.

\subsection{Graph community detection}

The final step in the analysis is to identify groups of SN spectra that are similar. We will start by looking at the simplified dissimilarity matrix shown in Figure \ref{fig:heatmap_mst}. To identify groups of similar objects, known as "communities," we will use the {\tt Infomap} algorithm \citep{Rosvall2008}.

The {\tt Infomap} algorithm is a method for detecting the community structure of networks. It is based on finding the most probable path a random walker would take through the network. The algorithm uses this information to identify groups of nodes (i.e. the communities) that are strongly connected within themselves but less connected to other groups. It does this by minimising the information cost of the paths taken by the random walker and using this to identify the most likely community structure of the network. The method is effective in identifying communities in a wide range of networks \citep{Lancichinetti2009}. The {\tt Infomap} algorithm is implemented in the {\tt igraph} package \citep{Gabor2006}, which provides a user-friendly interface for its application.

\section{Results and Discussion}
\label{sec:groups}

\begin{figure*}
    \centering
    \includegraphics[width=0.95\linewidth]{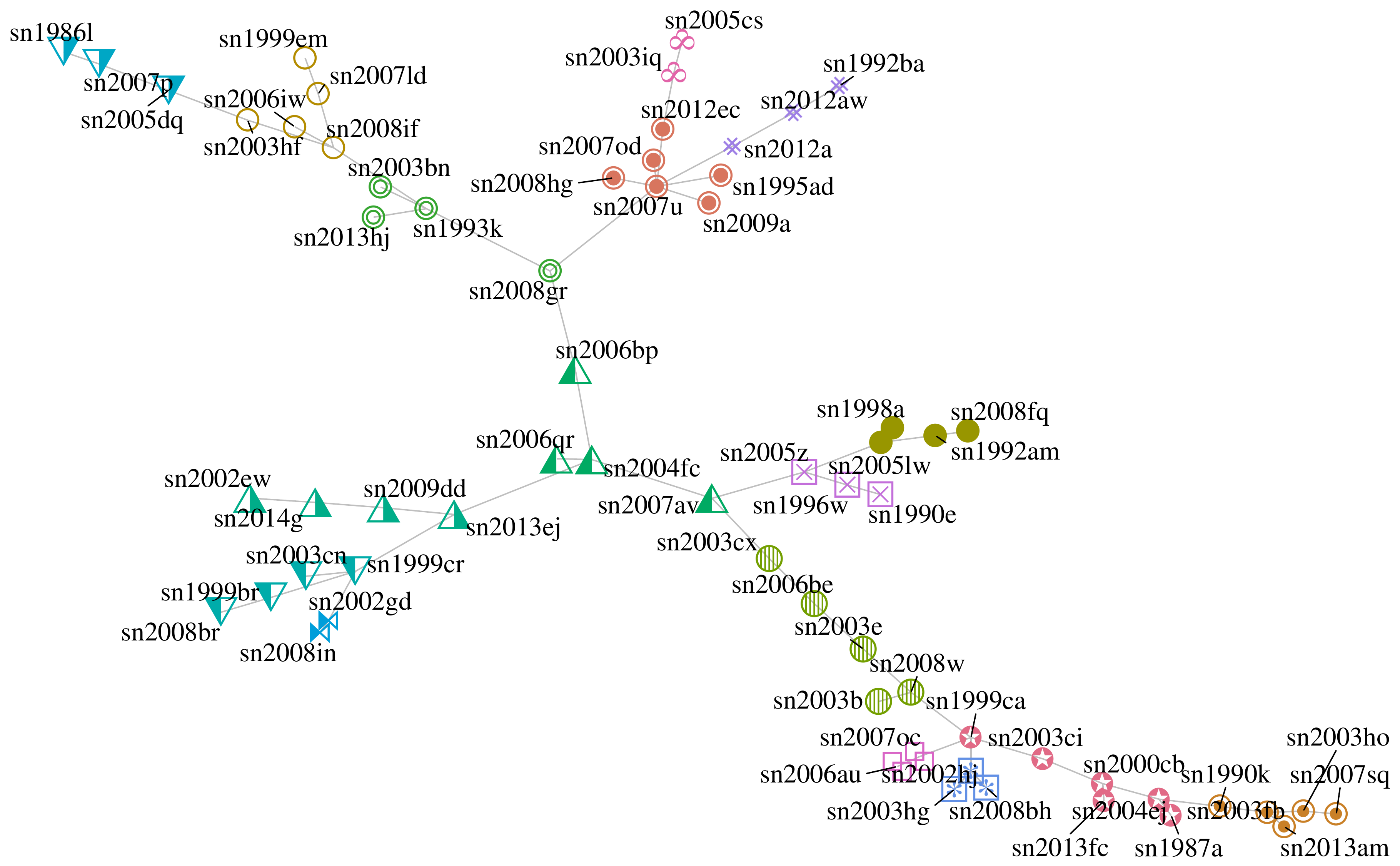}
    \includegraphics[width=0.95\linewidth]{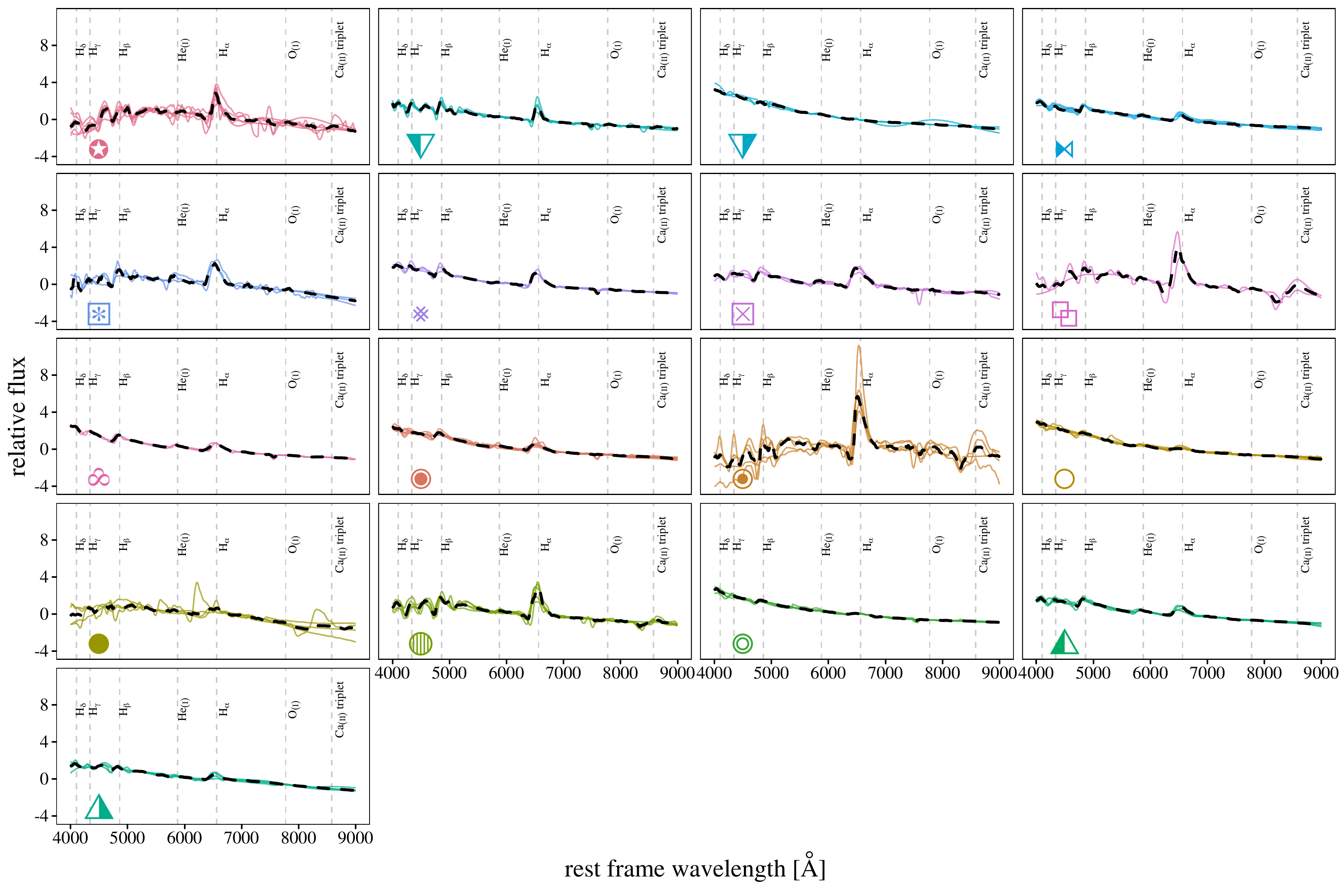}
    \caption{Top: Graph representation of SNe at maximum light. Bottom: Comparison of the diversity of spectra at maximum light in each group. The median spectrum is shown with a dashed black line.}
    \label{fig:graph_max}
\end{figure*}

\begin{figure*}
    \centering
     \includegraphics[width=0.95\linewidth]{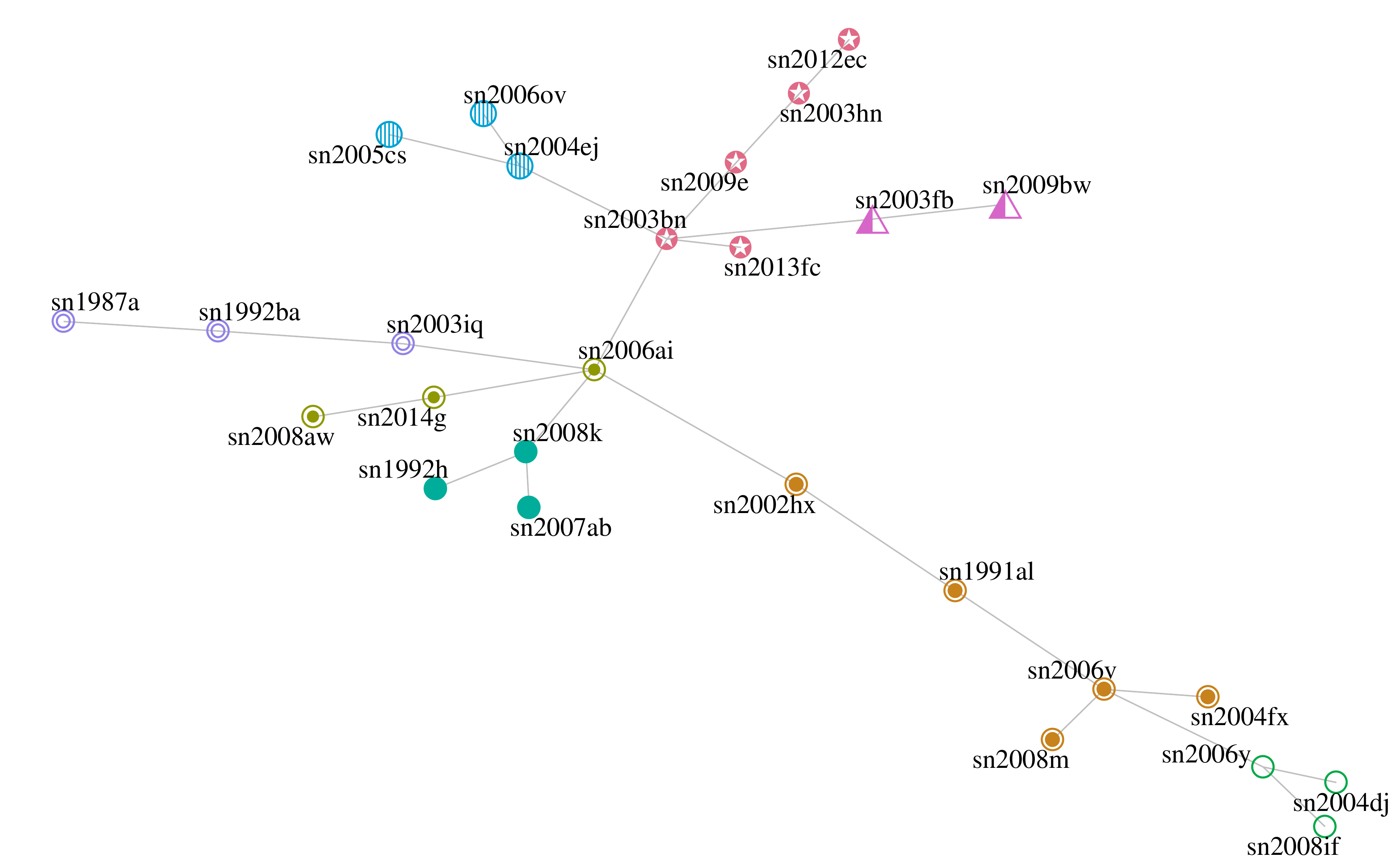}
         \includegraphics[width=0.95\linewidth]{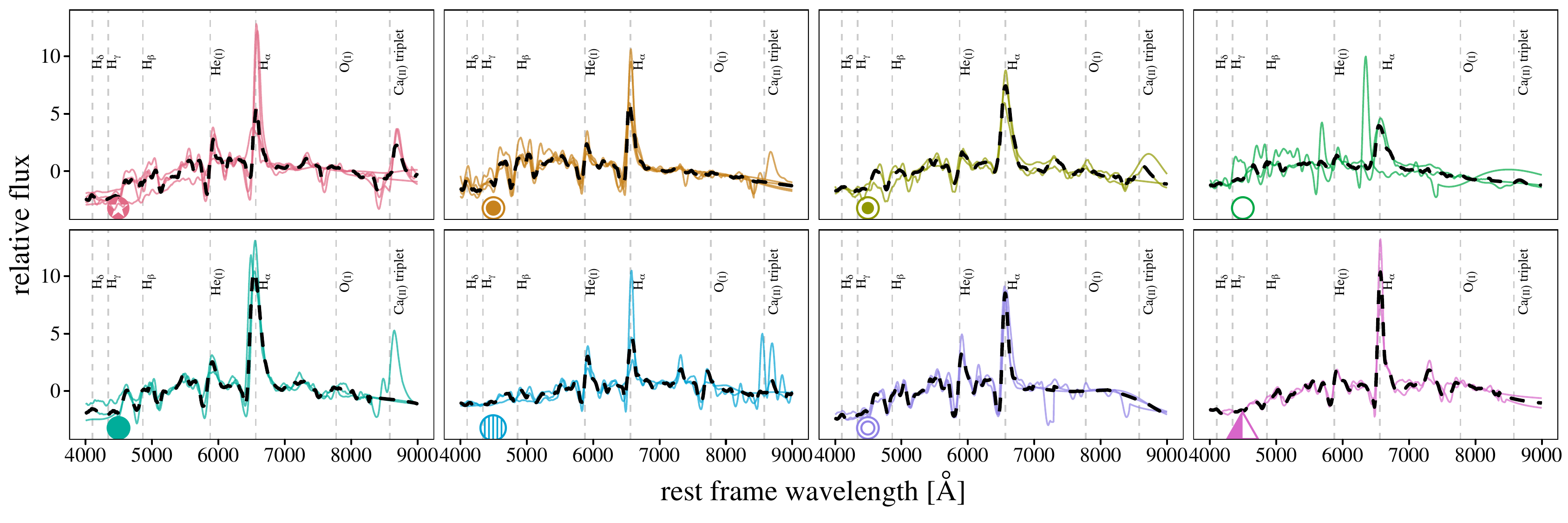}
    \caption{Top: Graph representation of SNe at end of the plateau phase. Bottom: Comparison of the diversity of spectra at end of the plateau phase in each group. The median spectrum is shown with a dashed black line.}
   \label{fig:graph_plateau}
\end{figure*}

The top panels of Figs.\ \ref{fig:graph_max} and \ref{fig:graph_plateau} present the graph network representation of the diversity of supernova (SN) spectra at maximum light and the end of the plateau phase, respectively. The bottom panels of these figures show the corresponding spectra in each group, along with the median spectra for each group (indicated by black dashed lines).

A visual inspection of these figures reveals several interesting features. The method used to construct the graph network representation effectively splits the spectra at maximum light based on their line features and the spectral slope at lower wavelengths. One of the key aspects that can be immediately recognized is the continuum of SNe from the top-right part of the graph towards the bottom-left, which traces the relative intensity of $\mathrm{H}_{\alpha}$. The spectra at the end of the plateau phase appear to show less diversity than those at maximum light, although the smaller sample size for this phase prevents us from making strong claims about the relative diversity of the spectra. The figures presented are a valuable tool for understanding the diversity of supernovae spectra. They enable a clear comparison of similarities and differences among different groups of spectra. They can be used to discover new SNe subclasses or examine the relationships between existing subclasses. 

\begin{figure*}
    \centering
    \includegraphics[width=0.95\linewidth]{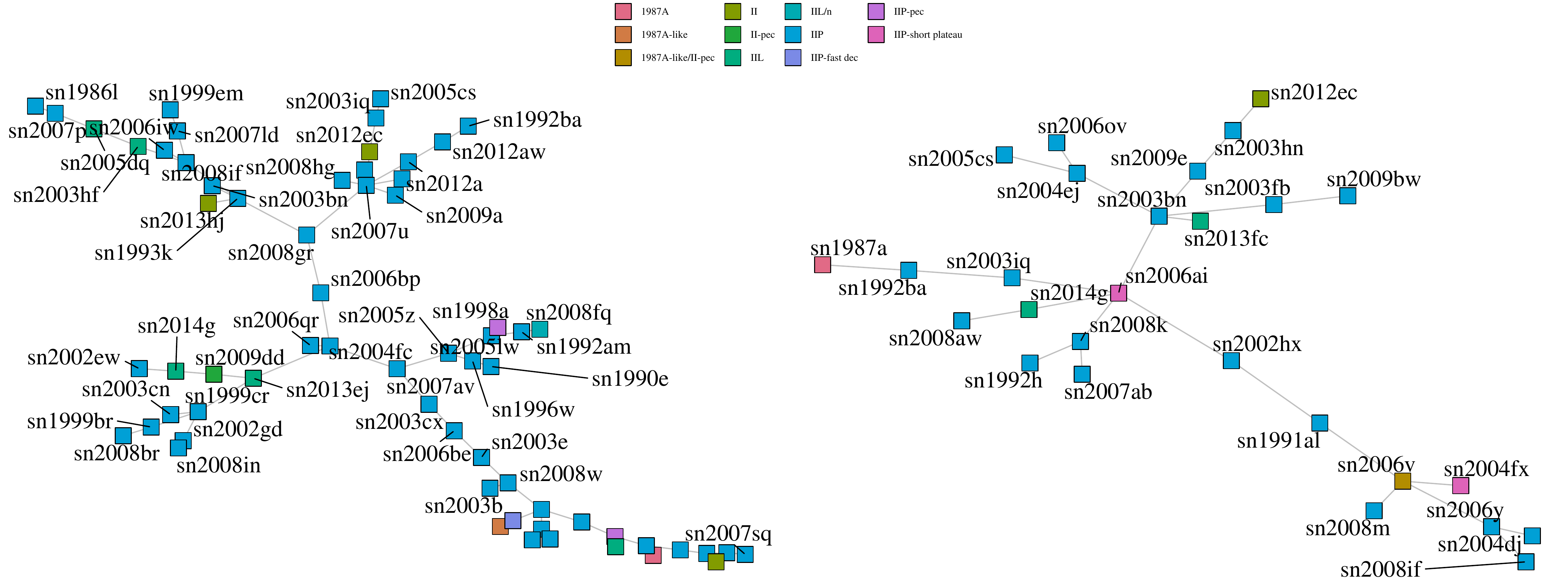}
    \caption{On the left graph, we have a representation of supernovae (SNe) at their maximum light. On the right graph, we represent SNe at the end of their plateau phase. In both graphs, the SNe are colour-coded based on their standard classification in the literature.}
    \label{fig:graph_type}
\end{figure*}

In the following, we compare  the results of our graph classification method with those of a more traditional classification method found in the literature. Additional insights can be derived from Figure \ref{fig:graph_type}.
SN2009dd is a Type II supernova that displays characteristics of both the bright branch and weak interaction between the circumstellar material and the ejecta, as indicated by high-velocity features in the Balmer lines \citep{Inserra2013}. This feature is commonly observed in Type IIn SNe and Type IIL SNe \citep{Bostroem2019}. Our graph-based analysis positions SN2009dd between SN2014g and SN2005lw, both Type IIL at maximum light phase, suggesting that SN2009dd may be a transitional link between standard Type IIP and strongly interacting Type IIn supernovae. This further highlights the challenges in devising a clear classification scheme for these objects, as luminous Type II SNe like SN2009dd, can exhibit properties of multiple supernova types.

 SN2006ai is noteworthy for its central position in the graph and its classification as a transitional type between Type IIL and IIb supernovae, with short-plateau phases, potentially indicating a high-mass red supergiant progenitor  \citep{Hiramatsu_2021}. Additionally, SN2014g, located close to SN2006ai, is classified as a Type IIL supernova, with a post-maximum light curve decline consistent with this classification. However, it was initially classified as a Type IIn due to strong emission lines in the earliest days of the explosion, likely caused by a metal-rich circumstellar medium (CSM), possibly resulting from pre-explosion mass loss events \citep{Terreran2016}. SN2008aw also shows an extra absorption component on the blue side of $H_\alpha$ \citep{Gutierrez2014}. Not surprisingly, SN1987A appears to deviate from the general trend on the graph and is located on the periphery, while still maintaining a connection to other Type IIP supernovae.

 While an in-depth examination of individual supernovae is not within the scope of this work, the examples above demonstrate how our approach can assist domain specialists in identifying objects within a population. These identified objects may provide crucial insights into their subsets' composition.

\begin{figure*}
    \centering
     \includegraphics[width=0.475\linewidth]{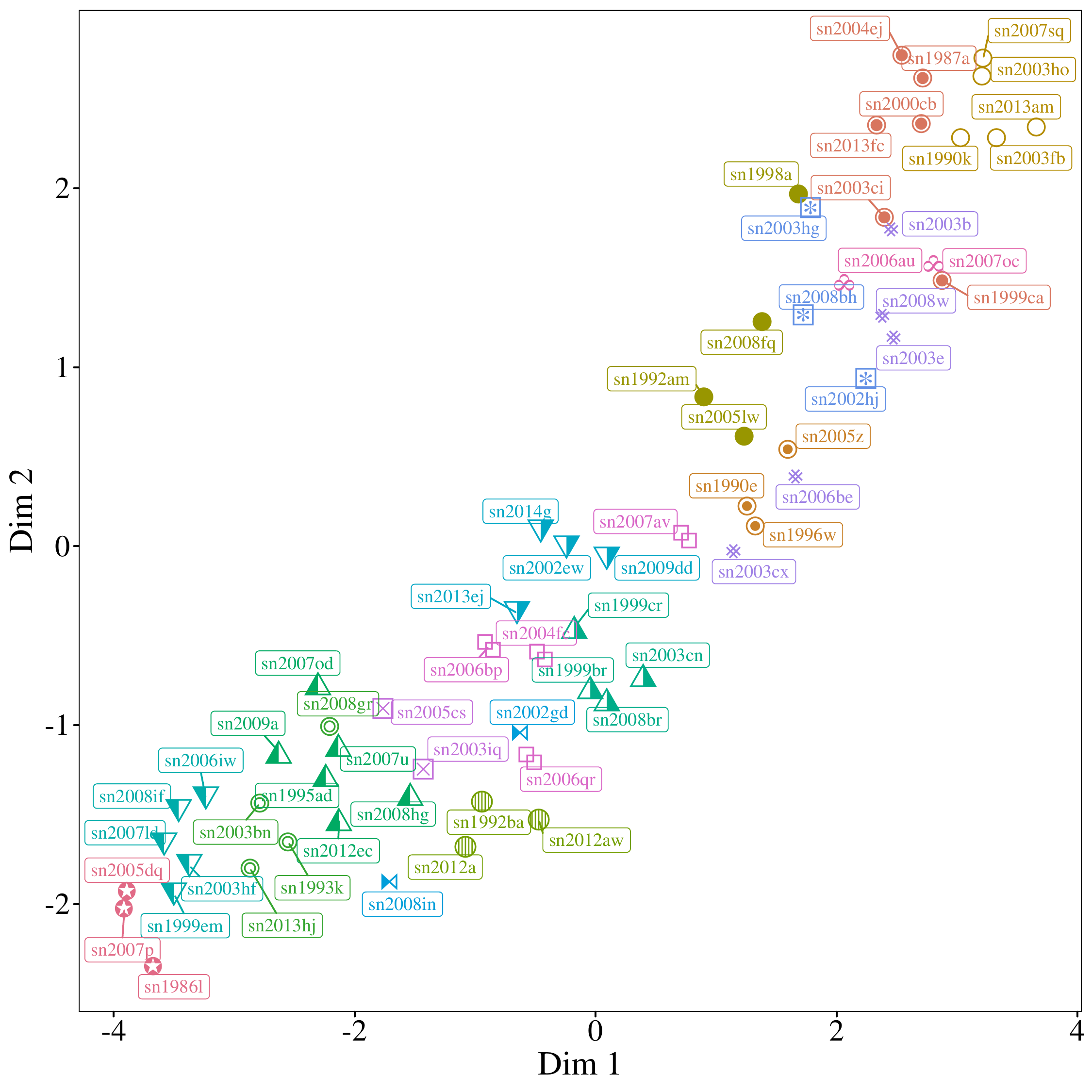}
     \includegraphics[width=0.475\linewidth]{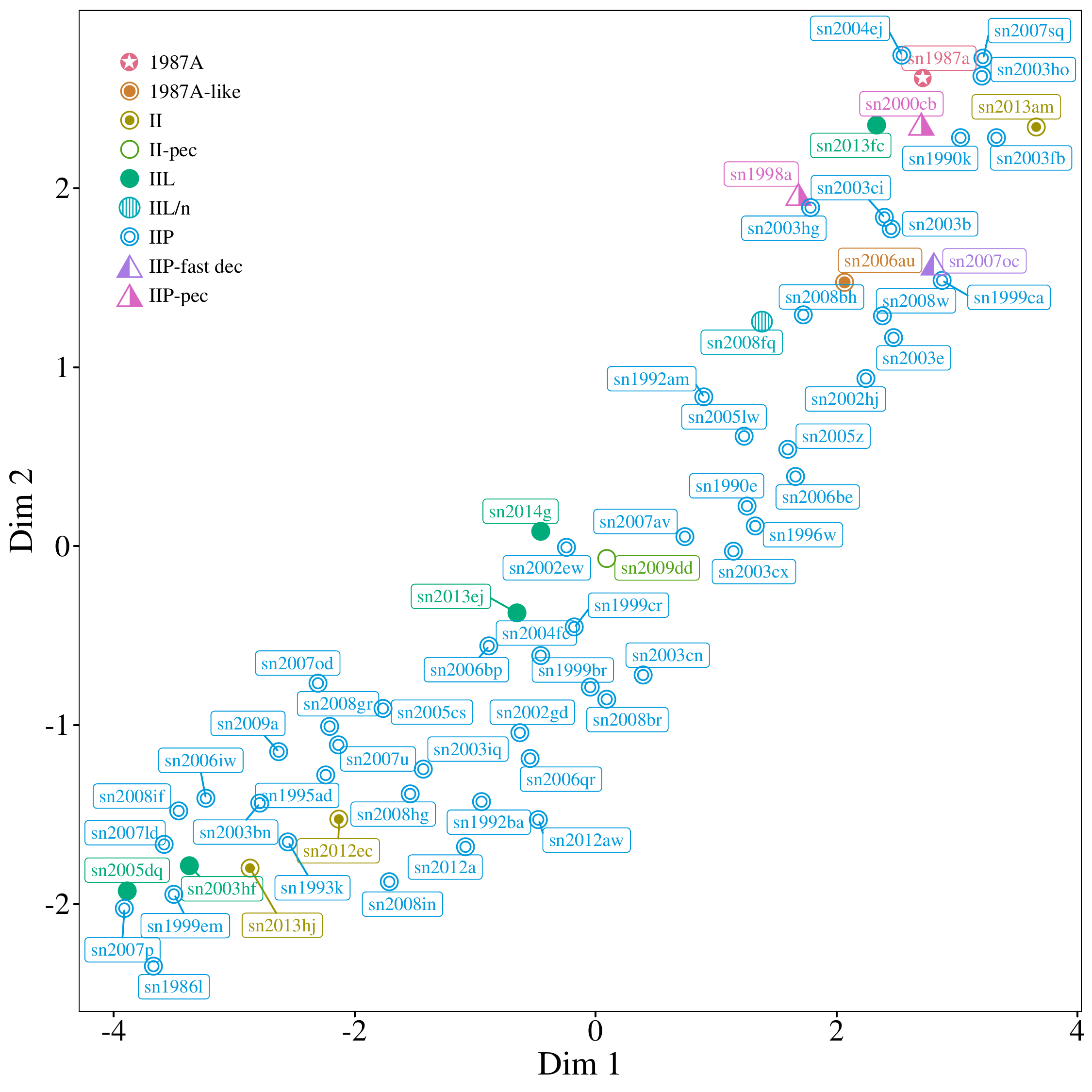}
      
    \caption{The two-dimensional embedding of SN spectra by \umap at maximum light is depicted in two panels. The left panel displays our graph-based community groups, while the right panel depicts the standard classification from the literature. In both panels, each SN is coded with a unique colour and shape according to its respective classification, which was assigned independently of the \umap embedding. }
   \label{fig:umap_max}
\end{figure*}

\begin{figure*}
    \centering 
         \includegraphics[width=0.475\linewidth]{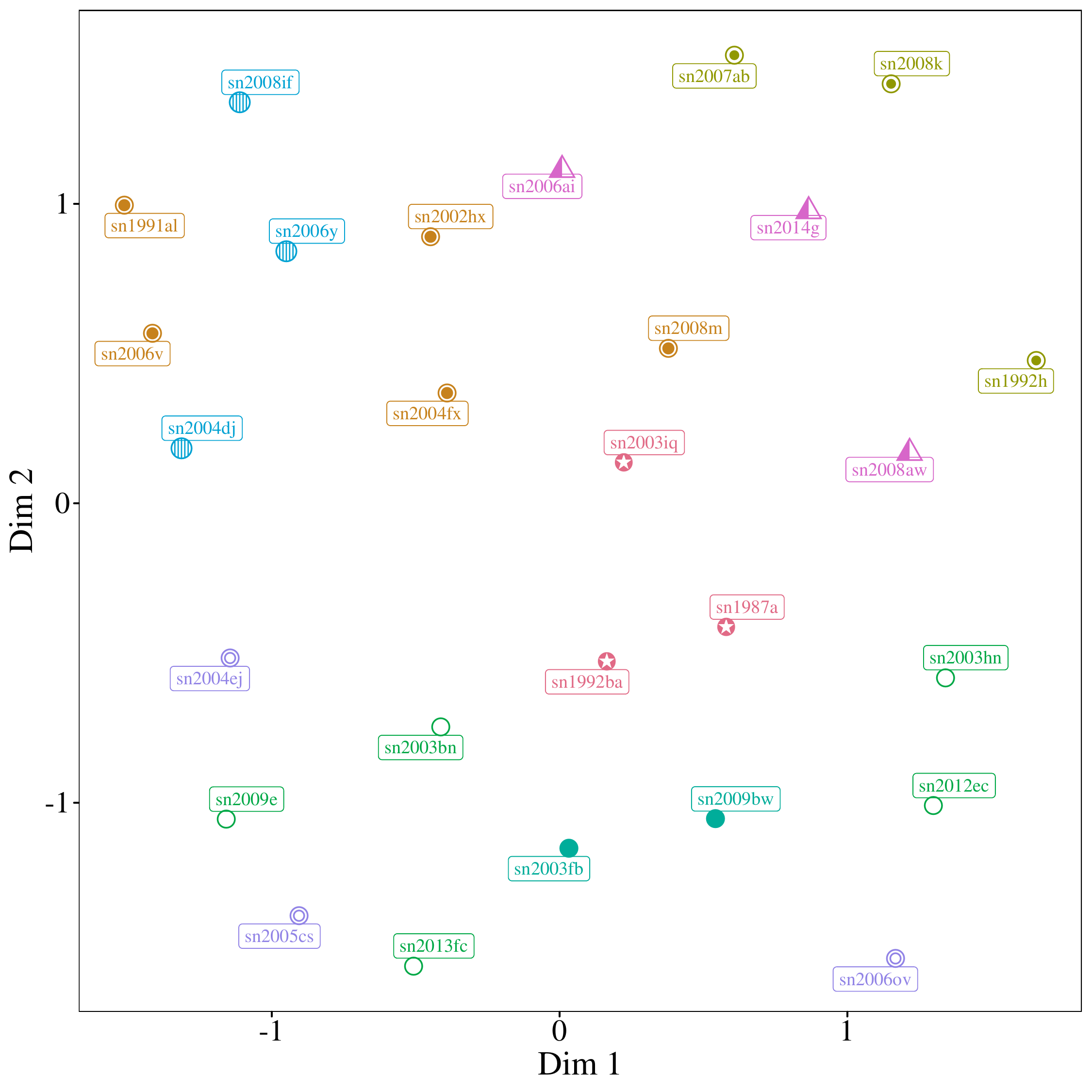}
            \includegraphics[width=0.475\linewidth]{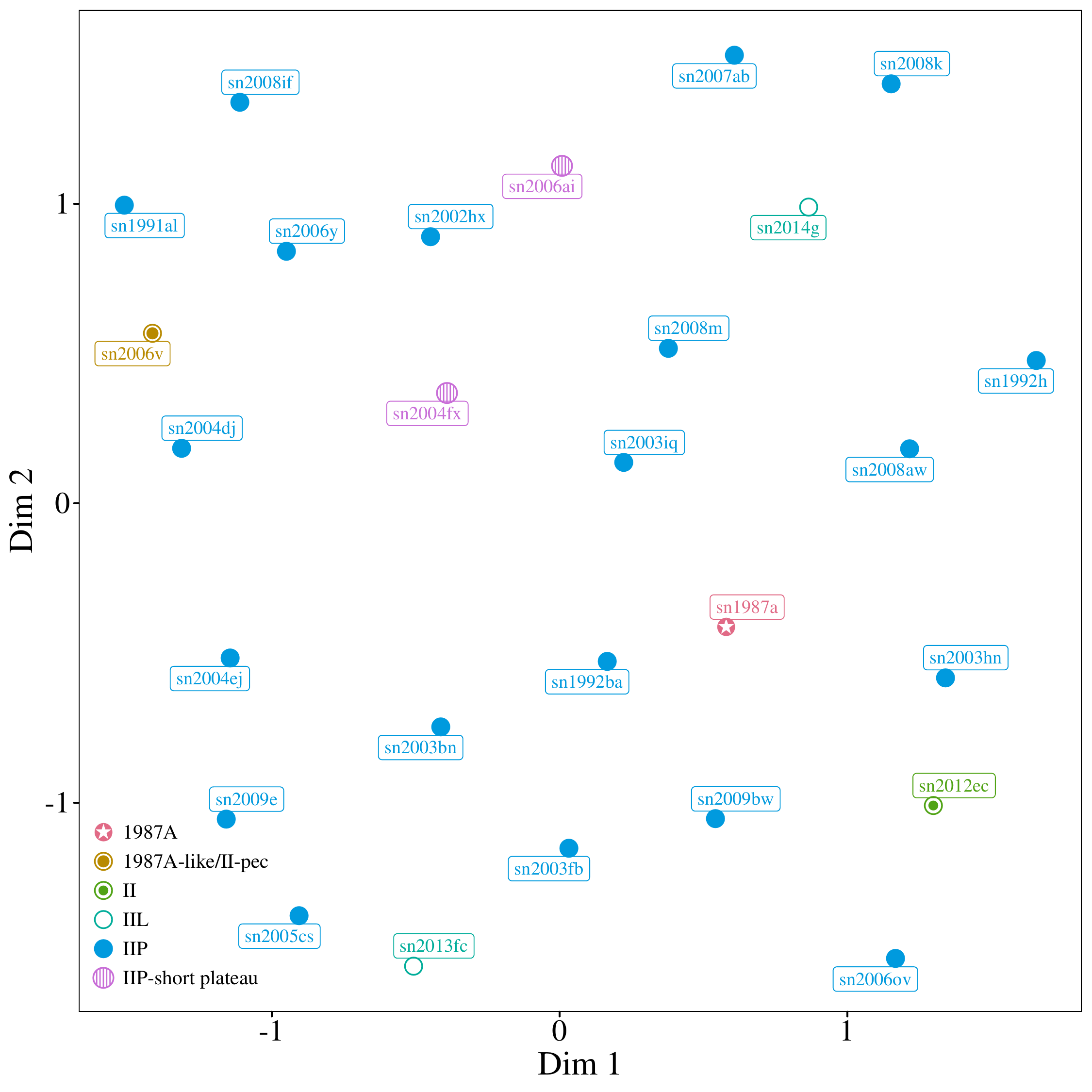}
    \caption{The two-dimensional embedding of SN spectra by \umap at the end of the plateau phase is depicted in two panels. The left panel displays our graph-based community groups, while the right panel depicts the standard classification from the literature. In both panels, each SN is coded with a unique colour and shape according to its respective classification, which was assigned independently of the \umap embedding.}
   \label{fig:umap_pt}
\end{figure*}

\subsection{UMAP projection}

For an independent visualisation of the SNe spectra at each phase, we use the Uniform Manifold Approximation and Projection ({\sc umap}) algorithm \citep{McInnes2018}. \umap is a manifold learning algorithm that aims to project high-dimensional data into a lower dimension while preserving local distances over global distances. It is similar in scope to other manifold learning algorithms such as t-SNE \citep{tsne}, ISOMAPs \citep{isomap}, and diffusion maps \citep{diffusions2005}. However, \umap is computationally more efficient and has been shown to have better discriminating power than other state-of-the-art methods such as t-SNE \citep{McInnes2018}. 

Figures \ref{fig:umap_max}, and \ref{fig:umap_pt} present a two-dimensional projection of supernova (SN) spectra, which are colour-coded and shape-coded according to the groups identified by graph analysis. A visual examination of the figure reveals a strong correspondence between these two independent methodologies. It is worth noting that the groups were not utilised in the \umap projection and are highlighted solely for visualisation purposes. A closer examination of the \umap projection confirms that while the SN spectra exhibit distinct features, particularly around maximum light, they form a continuum that ranges from the high blue slope, featureless spectra to spectra dominated by $\mathrm{H}_{\alpha}$ features. This figure provides independent validation of the results obtained through graph analysis and further supports the conclusion that the SN spectra constitute a continuous family, displaying varying spectral features.

Although the limited size of our dataset hinders our ability to make definitive statements, we were able to draw some anecdotal insights from the 11 supernova (SN) spectra present in both phases. Notably, sn2004ej and sn2013fc appear next to each other in the graphs and \umap representations, as do sn2003iq and sn1992ba. This similarity in positioning suggests a similar temporal evolution of these objects.
Furthermore, an Adjusted Rand Index (ARI) of 0.5\footnote{The Adjusted Rand Index (ARI) is a measure used to evaluate the similarity between two clustering solutions, such as the true clustering and a predicted clustering. It considers all possible pairs of items and compares their assignments in both solutions. ARI ranges from -1 to 1, where a score of 1 indicates a perfect match between the two clusterings, 0 means the similarity is no better than random chance, and negative values indicate disagreement.} was calculated for these 11 objects, indicating a moderate agreement between the group assignments. This result suggests some coherence in how SNe spectra evolve, but also indicates a certain degree of variance between different SNe spectra.

 Both traditional classification and our methodology reveal distinct patterns in the \umap space. However, upon visual inspection of the \umap projection, it is apparent that our classification scheme presents a more unified and seamless distribution of classes solely based on SN spectra. Although this analysis cannot infer any potential differences in the progenitor, our methodology offers a more quantitative approach to organising SNe based on their spectra alone. Additionally, it demonstrates consistency with more advanced and independent non-linear feature extraction analysis. The motivation behind this step is to provide a validation check of our previous analysis, which involved a series of straightforward steps, from the computation of the dissimilarity matrix to the construction of the network. 

 \umap, on the other hand, is a non-linear dimensionality reduction technique. The consistency between these two methodologies suggests that a small set of primary properties can explain the SN spectra.

Similarly to any data-driven approach, the results presented in this work are bounded by the amount of information contained in the initial data set. Nevertheless, within the constraints imposed by the available data, the results presented here are a first hint of the potential application of such techniques. A larger and more diverse data set would help to verify or refute the relationships suggested here and possibly enable new discoveries.

\section{Conclusions}
\label{sec:concl}

This study presents a novel heuristic approach for automated spectral classification, which combines pair-wise $\ell_1$-norm dissimilarity and graph community detection. The main advantage of this method is its ability to create a continuous representation of the spectral taxonomy, suitable for capturing major groups, outliers, and transitional types. We have applied our method to a sample of supernova spectra from various catalogues and analyzed it in two spectral phases: maximum light and end of the plateau phase.

Around maximum light, our method effectively captures the fast evolution of Type II supernova spectra, translating it into a structured graph. Additionally, it highlights their similarities and enables their representation as a continuum ranging from featureless and continuum-dominated to $\mathrm{H}_{\alpha}$ dominated groups. On the other hand, the spectra around the end of the plateau phase give rise to a less complex graph structure, revealing the more extensive homogeneity among the members of this group.

Independent analysis using manifold learning projection confirms this result. This methodology is versatile and should be readily applicable to other unsupervised spectral classification problems or, more broadly, to other types of functional data in astronomy, such as photometric time series. The consistency between the two completely distinct analyses (graph-based clustering and \umap) and the ability to plug in a physically motivated distance metric between spectra is a plus.  This approach represents a more flexible paradigm than traditionally employed classification schemes.  A snippet code to map a tabular data into a network visualization is publicly available within the COINtoolbox\footnote{\url{https://github.com/COINtoolbox/graph_clustering}}.

Overall, this study presents a powerful new tool for the analysis of spectral data, which may help to enhance our  understanding of the underlying physical mechanisms driving the observed phenomena.

\section*{Acknowledgements}

The authors thank Christian Vogl, Wolfgang Kerzendorf, and Jordi Nadal for their fruitful discussions and suggestions.
This work is a result of the $\rm 6^{th}$ COIN Residence Program\footnote{\url{https://cosmostatistics-initiative.org/residence-programs/crp6}} (CRP\#6) held in Chamonix, France in August 2019. COIN was financially supported by CNRS as part of its MOMENTUM programme over the 2018 -- 2020 period.
R.S.S.\ was supported by the National Natural Science Foundation of China project 1201101284.
S.T.\ was supported by the Cambridge Centre for Doctoral Training in Data-Intensive Science, funded by the U.K. Science and Technology Facilities Council (STFC).
S.G.G.\ acknowledges support by the FCT under Project CRISP PTDC/FIS-AST-31546/2017 and Project No.\ UIDB/00099/2020.
L.G.\ acknowledges financial support from the Spanish Ministerio de Ciencia e Innovaci\'on (MCIN), the Agencia Estatal de Investigaci\'on (AEI) 10.13039/501100011033, the European Social Fund (ESF) "Investing in your future" under the 2019 Ram\'on y Cajal program RYC2019-027683-I and the PID2020-115253GA-I00 HOSTFLOWS project, from Centro Superior de Investigaciones Cient\'ificas (CSIC) under the PIE project 20215AT016, and the program Unidad de Excelencia Mar\'ia de Maeztu CEX2020-001058-M.  The Cosmostatistics Initiative\footnote{\url{https://cosmostatistics-initiative.org/}} (COIN) is an international network of researchers whose goal is to foster interdisciplinarity inspired by Astronomy.




\bibliography{ref} 



\appendix
\section{Impact of $\ell$-norm selection}
\label{app:lnorm}

\begin{figure*}
    \centering
    \includegraphics[width=0.45\linewidth]{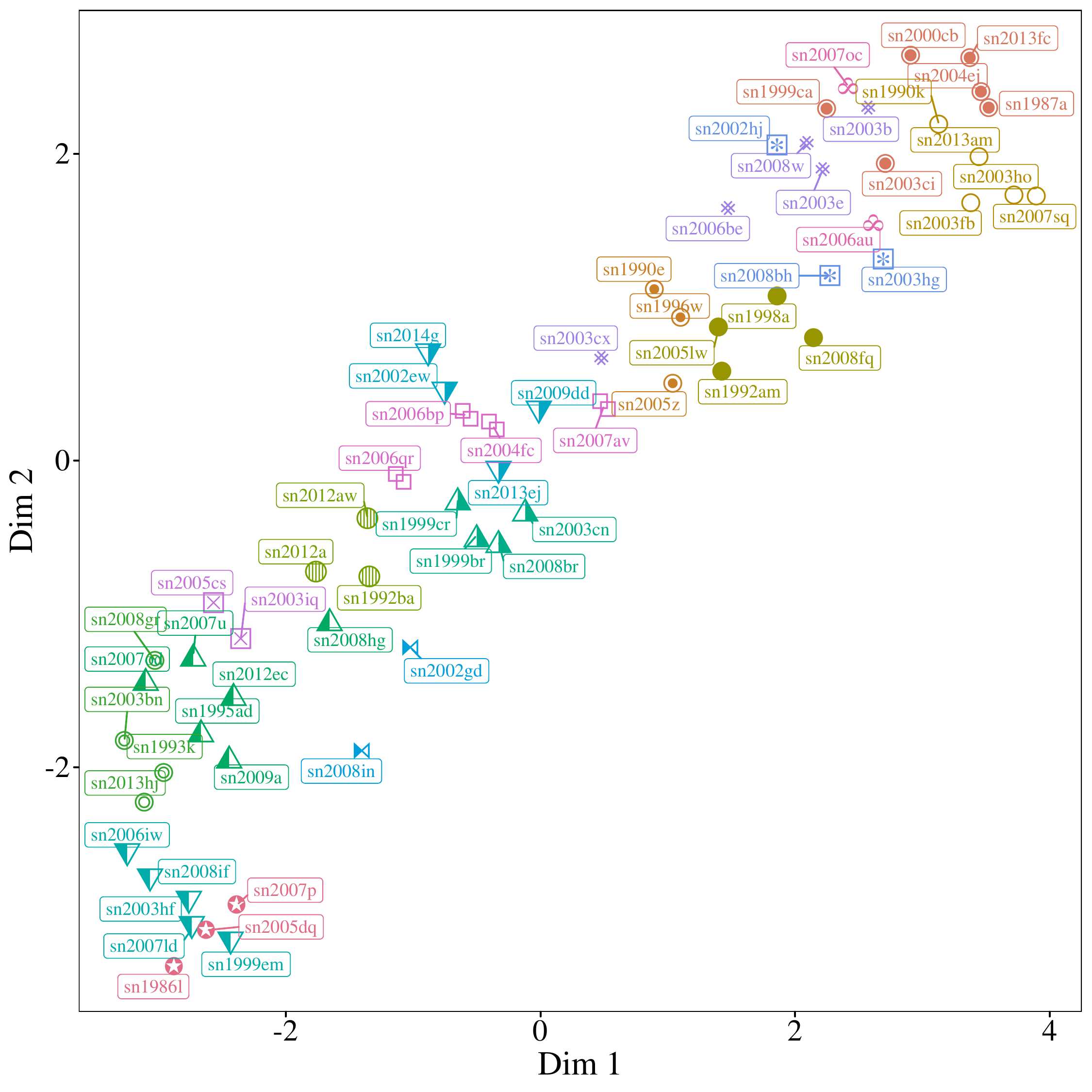}
    \includegraphics[width=0.45\linewidth]{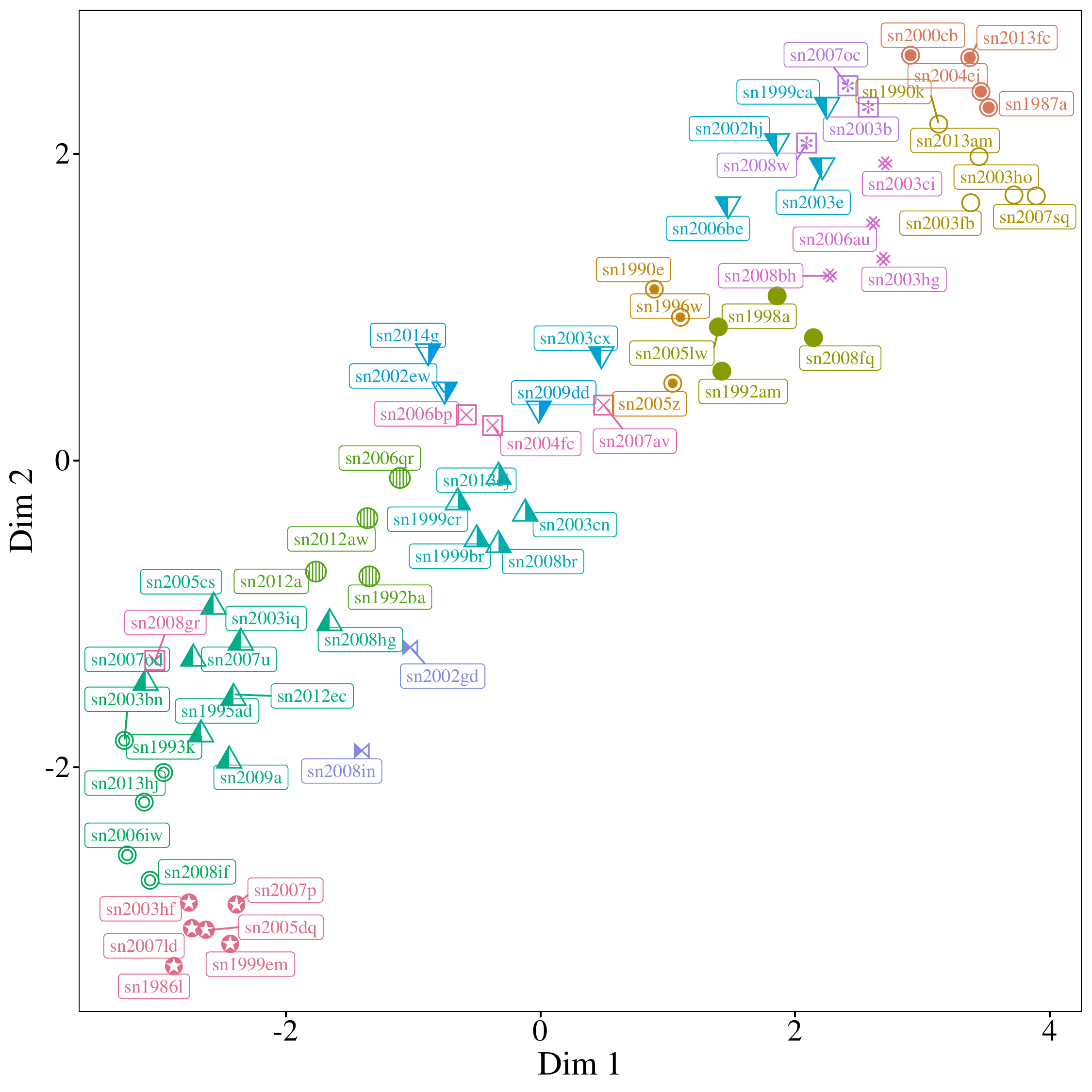}
    \caption{The visualisation of SN spectra through a two-dimensional embedding by \umap is presented in two panels. The left panel shows the graph-based community groups using Manhattan distance, while the right panel showcases the classification based on Euclidean distance between the SNe spectra. Each SN is distinguished with a unique colour and shape corresponding to its classification on each panel.   Note, however, that the specific color and shape assigned to each cluster may vary between the panels.}
     \label{fig:graph_maxL2}
   \end{figure*}

\begin{figure*}
    \centering
    \includegraphics[width=0.45\linewidth]{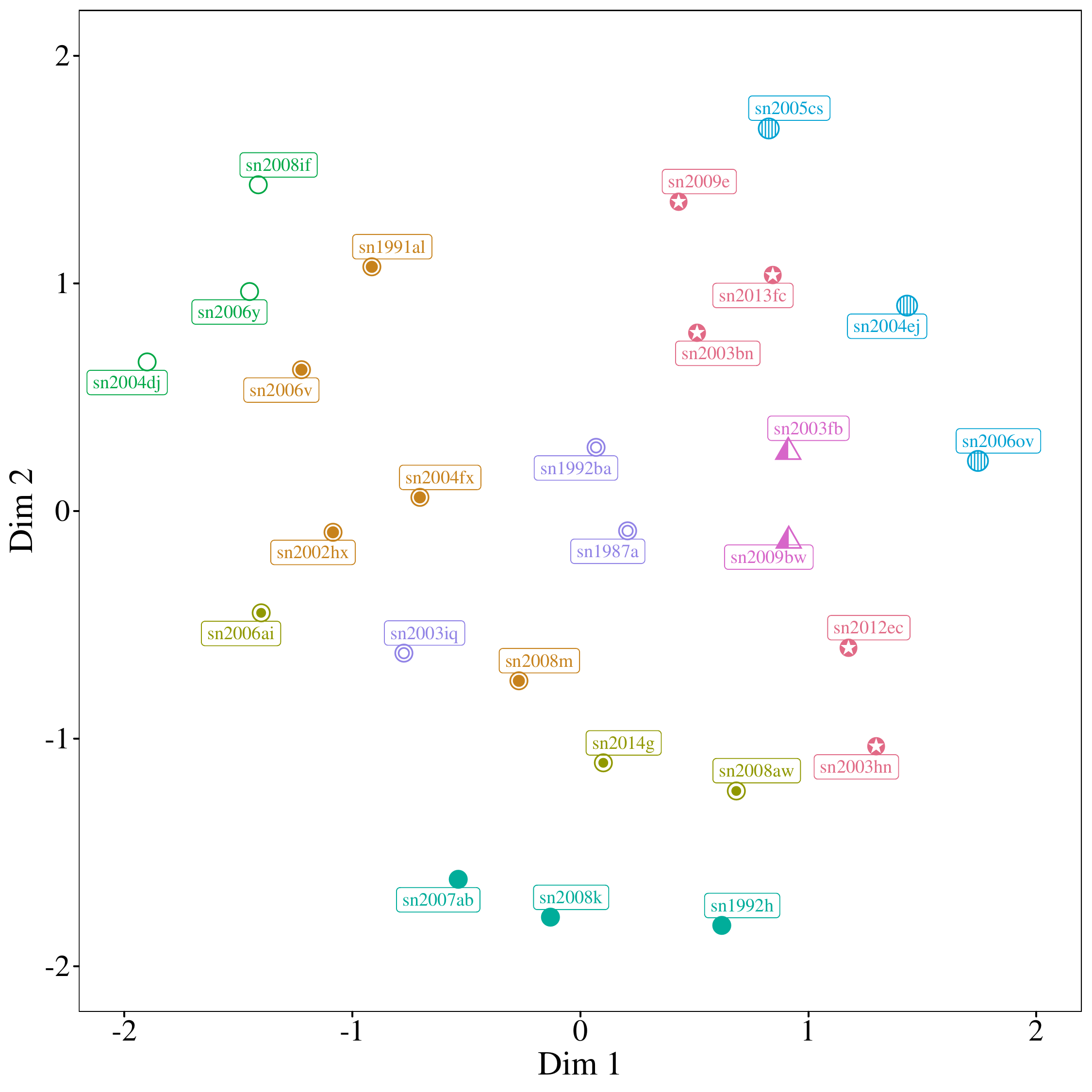}
    \includegraphics[width=0.45\linewidth]{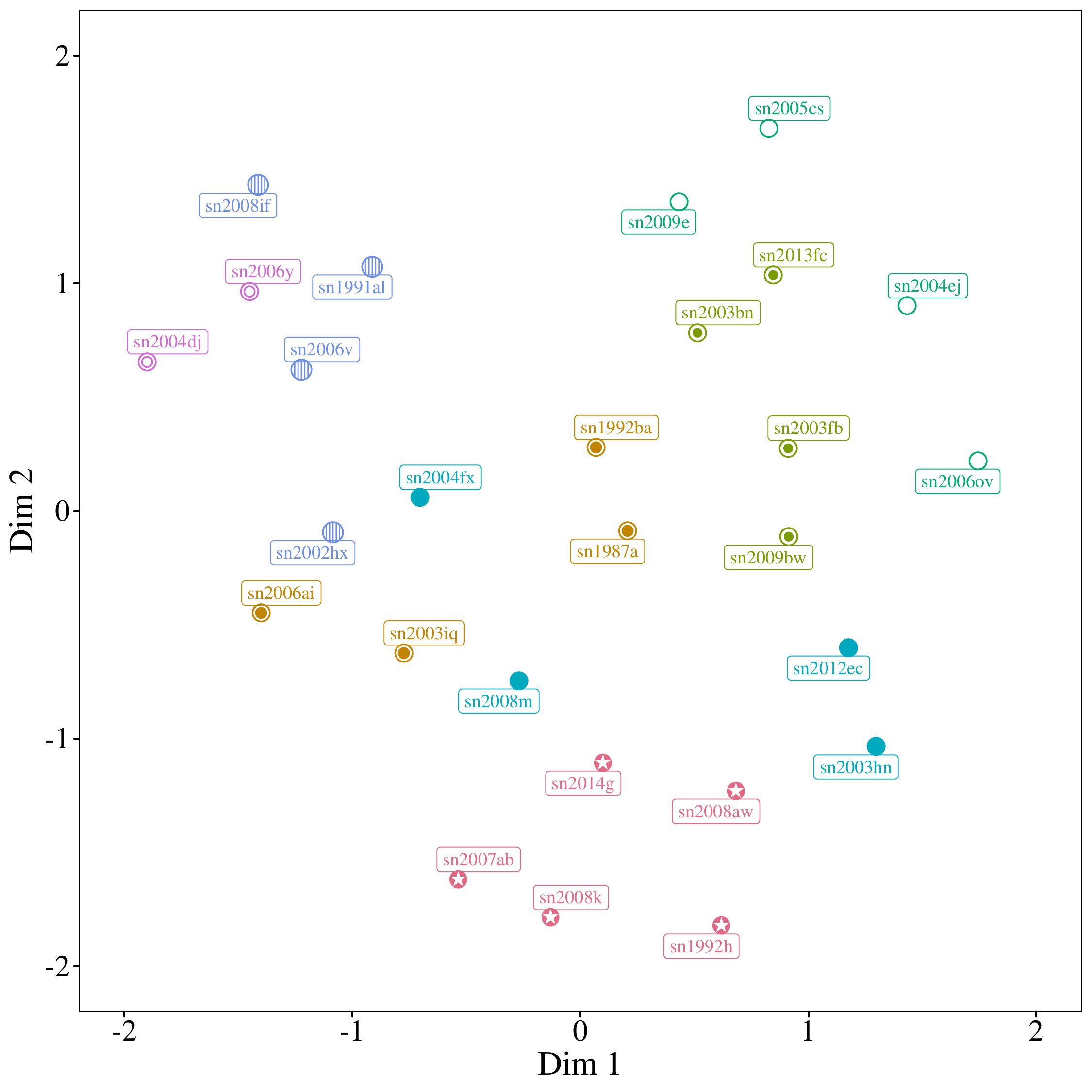}
    \caption{The visualisation of SN spectra through a two-dimensional embedding by \umap is presented in two panels. The left panel shows the graph-based community groups using Manhattan distance, while the right panel showcases the classification based on Euclidean distance between the SNe spectra. Each SN is distinguished by a unique color and shape corresponding to its classification in both panels.  Note, however, that the specific color and shape assigned to each cluster may vary between the panels.}
     \label{fig:graph_ptL2}
   \end{figure*}

In this appendix, we examine the impact of different $\ell$-norm choices on the performance of our graph-based supernovae classification approach. Our approach remains unchanged, except for estimating the dissimilarity matrix, which is based on the chosen $\ell$-norm. The results of using a $\ell_2$-norm (Euclidean), in comparison to $\ell_1$-norm (Manhattan), are shown in Figures \ref{fig:graph_maxL2} and \ref{fig:graph_ptL2} at maximum light and end of plateau phase respectively.  The number of assigned groups varies with the choice of metric, with 15 groups found using $\ell_2$-norm and 17 groups found using 
 $\ell_1$-norm. This results in differences in the shapes and colours across panels.
The visual examination shows that despite some differences, the group structure remains consistent in the \umap projection and spectral feature variations. For instance, using the Euclidean norm combines SN19861 and SN1999em into one group, while the $\ell_1$-norm splits the groups (\autoref{fig:graph_maxL2}). During the end of the plateau phase, the membership assignment is not heavily affected by the metric choice, but there is some shuffling due to a lack of a hard cut to separate SNe by their spectra (\autoref{fig:graph_ptL2}). The $\ell_2$-norm emphasises the overall spectral shape, while the $\ell_1$-norm is more sensitive to sparse features like spectral lines. Yet, our results suggest that the observed structure in the data is not an artefact of the specific distance measure used in our analysis.

\end{document}